\documentclass[aps,pra,epsfig]{revtex4}
\usepackage{graphicx}
\usepackage{epsfig}
\usepackage{amssymb}
\usepackage{graphicx}
\usepackage{graphics}
\usepackage{amssymb}
\usepackage{amsmath}
\usepackage{epsfig}
\usepackage{latexsym}
\usepackage{color}

\begin{document}
\title{Quasiprobability methods for multimode conditional optical gates.}
\author{G.J. Milburn}
\address{Centre for Quantum Computer Technology,
The University of Queensland,
QLD 4072, Australia.}
\email{milburn@physics.uq.edu.au}

\begin{abstract}
We present a method for computing the action of conditional linear optical transformations, conditioned on photon counting, for arbitrary signal states. The method is based on the Q-function, a quasi probability distribution for anti normally ordered moments. We treat an arbitrary number of signal and ancilla modes. The ancilla modes are prepared in an arbitrary product number state. We construct the conditional, non unitary, signal transformations for an arbitrary photon number count on each of the ancilla modes. 
 \end{abstract}
\maketitle
It has been known for some time  that non deterministic non linear transformations are possible with linear optical networks when some subset of the  input modes (ancillas) are prepared in single photon states before the optical network and directed to photon counters at the network output\cite{KLM,Kok,Scheel1,Eisert,Rudolph}. The conditional state of all non-ancilla modes (the {\em signal} modes), conditioned on a particular count on the output ancilla modes, is given by a non unitary transformation of the input signal state and can simulate a highly nonlinear optical process. This transformation is defined in terms of a conditional measurement operator acting on the signal modes alone. It is determined entirely by the total linear optical unitary transformation (acting on both signal and ancilla modes), the known ancilla input number states and the ancilla photon counts at the output. A number of approaches have been proposed for constructing the conditional state transformation.  Scheel et al.\cite{Scheel2} give an algebraic approach, while Lapaire et al.\cite{Lapaire} show how nonlinear such transformations can be. In \cite{vanLoock} van Loock et al. derive a set of criteria to decide whether a given measurement operator can be implemented.   In this paper we show how quasi probability methods (the Q function) can be used to construct the measurement operator for arbitrary ancilla number states and arbitrary ancilla number counts.  The method leads naturally to a diagrammatic representation of the measurement operator. 

\begin{figure}[h!]
\includegraphics[scale=0.6]{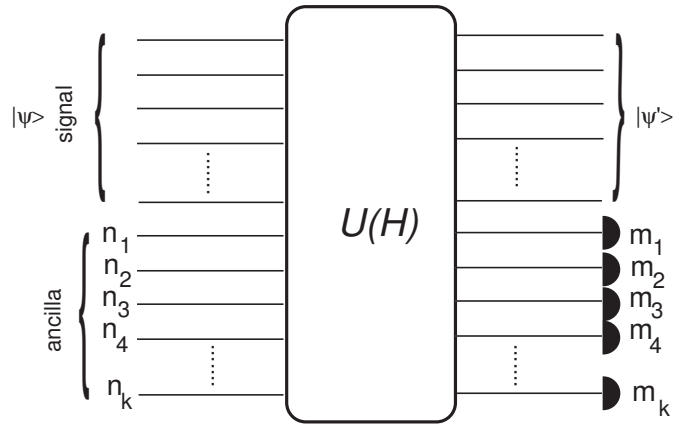}
\caption{A conditional linear optical gate.} \label{fig1}
\end{figure}

Consider the situation depicted in figure \ref{fig1}. in this device $N+K$ modes pass through a optical device that conserves total photon number. We will call this a linear optical system. The $K$ ancilla modes are prepared in photon number eigenstates. At the output, photon number measurements are made on the ancilla modes alone.  We seek the {\em conditional} state for the remaining $N$ modes {\em given} the ancilla photon number count.

The linear optical device performs a unitary transformation on all the input states. It may be described by a unitary transformation of the form\cite{Scheel2,Aniello}
\begin{equation}
U(H)=\exp[-i\vec{a}^\dagger H\vec{a}]
\end{equation}
where 
\begin{equation}
\vec{a}=\left (\begin{array}{c}
				a_1\\
				a_2\\
				\vdots\\
				a_N\\
				a_{N+1}\\
				\vdots\\
				a_{N+K}
				\end{array}\right )
\end{equation}
and $H$ is a hermitian matrix. This transformation leaves the total photon number invariant,
\begin{equation}
U^\dagger(H)\vec{a}^\dagger\cdot\vec{a}U(H)=\vec{a}^\dagger\cdot\vec{a}
\end{equation}
It also induces a linear unitary transformation on the vector $\vec{a}$ as
\begin{equation}
U^\dagger(H)\vec{a}U(H)=S(H)\vec{a}
\end{equation}
One should not confuse the unitary transformation, $U(H)$, acting on states, with the induced unitary representation, $S(H)$, acting on the mode operators. 

The {\em conditional state} of the signal modes, $|\psi'\rangle$ is then determined by
\begin{equation}
|\psi'\rangle_s=\frac{1}{\sqrt{p(\vec{m})}}\hat{E}(\vec{n}|\vec{m})|\psi\rangle_s
\label{cond}
\end{equation}
where the observed count is represented by the vector of values $\vec{m}$, and the probability for this event is $p(\vec{m})$. The measurement operator is
\begin{equation}
 \hat{E}(\vec{n}|\vec{m})=\mbox{}_{anc}\langle \vec{m}|U(H)|\vec{n}\rangle_{anc}
 \end{equation}
 with 
 \begin{equation}
 |\vec{m}\rangle_{anc}=|m_1\rangle_{N+1}\otimes|m_2\rangle_{n+2}\otimes\ldots\otimes|m_k\rangle_{N+K}
 \end{equation}

In what follows we will need to know how coherent states transform under $U(H)$. We define
\begin{equation}
|\vec{\alpha}\rangle=|\alpha_1\rangle\otimes|\alpha_2\rangle\otimes\ldots|\alpha_{N+K}\rangle
\label{coh-vec}
\end{equation}
where 
\begin{equation}
\vec{\alpha}=\left (\begin{array}{c}
				\alpha_1\\
				\alpha_2\\
				\vdots\\
				\alpha_N\\
				\alpha_{N+1}\\
				\vdots\\
				\alpha_{N+K}
				\end{array}\right )
\end{equation}
Then
\begin{equation}
U(H)|\vec{\alpha}\rangle=|\vec{\alpha}(H)\rangle= |\alpha_1(H)\rangle\otimes|\alpha_2(H)\rangle\ldots|\alpha_{N+K}(H)\rangle
\label{trans-coh}
\end{equation}
where $\vec{\alpha}(H)=S^\dagger(H)\vec{\alpha}$ , reflecting the well known fact that coherent states do not become entangled through a linear optical transformation of the kind considered here.  

To prove this result, we work with the Bargmann coherent states\cite{Bargmann}, 
\begin{equation}
||\vec{\alpha}\rangle=e^{\vec{a}^\dagger.\vec{\alpha}}|0\rangle
\end{equation}
which are related to Glauber coherent states by $|\vec{\alpha}\rangle=\exp[-\vec{\alpha}^\dagger.\vec{\alpha}/2]||\vec{\alpha}\rangle$. 
Then
\begin{eqnarray*}
U(H)||\vec{\alpha}\rangle & = & U(H)e^{\vec{a}^\dagger.\vec{\alpha}}|0\rangle\\
					& = &  U(H)e^{\vec{a}^\dagger.\vec{\alpha}}U^\dagger(H)|0\rangle\\
					 &= & e^{\vec{a}^\dagger.S^\dagger(H)\vec{\alpha}}|0\rangle\\
					  &= & e^{\vec{a}^\dagger.\vec{\alpha}(H)}|0\rangle\\
					  & = & ||\vec{\alpha}(H)\rangle
 \end{eqnarray*}
 where $\vec{\alpha}(H)=S^\dagger(H)\vec{\alpha}$. Now as $\vec{\alpha}^\dagger(H).\vec{\alpha}(H)=\vec{\alpha}^\dagger.\vec{\alpha}$, we can write $U(H)|\vec{\alpha}\rangle=|\vec{\alpha}(H)\rangle$. 
 
 We now define the function 
 \begin{equation}
 {\cal U}_H(\vec{\alpha})=\langle \vec{\alpha}|U(H)|\vec{\alpha}\rangle=\langle \vec{\alpha}||U(H)||\vec{\alpha}\rangle e^{-\vec{\alpha}^\dagger\cdot\vec{\alpha}}
 \end{equation}
which completely determines the operator $U(H)$ (which acts in the full tensor product Hilbert space for both signal and ancilla modes, as we now show.
 
Due to the over completeness of the coherent state basis, this diagonal matrix element suffices to completely determine the operator\cite{Qsymbol}. To see this we note that if any operator is put in normal order 
\begin{equation}
\hat{A}=\sum_{n,m}a_{nm}(a^\dagger )^n a^m
\end{equation}
the Q-symbol is defined by 
\begin{equation}
{\cal A}(\alpha)=\langle \alpha|\hat{A}|\alpha\rangle
\end{equation}
As $a|\alpha\rangle=\alpha|\alpha\rangle$ we see that
\begin{equation}
{\cal A}(\alpha)=\sum_{n,m}a_{nm}(\alpha^*)^n\alpha^m
\end{equation}.

It is easy to see that the Q-symbol for $U(H)$ is given by 
 \begin{equation}
 {\cal U}_H(\vec{\alpha})=\exp[\vec{\alpha}^\dagger S(H)\vec{\alpha}]\exp{[-\vec{\alpha}^\dagger\cdot\vec{\alpha}]}
 \end{equation}
The Q-symbol, ${\cal A}(\vec{\alpha})$ completely determines the corresponding operator, $\hat{A}$, by the following method. 
\begin{itemize}
\item Expand ${\cal A}(\vec{\alpha})$ as a power series in $(\alpha^*_j)^n\alpha_k^m$
\item Order all conjugates to the left and make the replacements $(\alpha^*_j)^n\mapsto (a^\dagger_j)^n$ and
$\alpha_k^m\mapsto a_j^n$
\end{itemize}
This mapping between Q-symbols and operators can always be done efficiently with respect to the number of modes, and can thus be efficiently implemented in a symbolic computer package.  
 
 We can also define the Q-symbol for the measurement operator in Eq.(\ref{cond}) as
 \begin{equation}
 {\cal E}_H(\vec{\alpha}_s|\vec{n},\vec{m})=\mbox{}_s\langle \vec{\alpha}_s|\hat{E}(\vec{n}|\vec{m})|\vec{\alpha}_s\rangle_s
 \end{equation}
 where 
 \begin{equation}
\vec{\alpha}_s=\left (\begin{array}{c}
				\alpha_1\\
				\alpha_2\\
				\vdots\\
				\alpha_n
				\end{array}\right )
\end{equation}
where the subscript $s$ refers to the signal modes alone. 

Noting that
\begin{equation}
 |n\rangle=\left .\frac{1}{\sqrt{n!}}\left (\frac{\partial}{\partial \alpha}\right )^n ||\alpha\rangle\right |_{\alpha=0}
\end{equation}
and that the partial derivatives commute with $U(H)$,
it is easy to see that
\begin{eqnarray}
 {\cal E}_H(\vec{\alpha}_s|\vec{n},\vec{m}) & = & \left [m_{N+1}!\ldots m_{N+K}! n_{N+1}!\ldots n_{N+K}!\right ]^{-1/2}\left [\left (\partial^*_{N+1}\right )^{m_{N+1}}\ldots \left (\partial^*_{N+K}\right )^{m_{N+K}}\right . \\ \nonumber
 & & \times \left . \left (\partial_{N+1}\right )^{n_{N+1}}\ldots \left (\partial_{N+K}\right )^{n_{N+K}}\exp[\vec{\alpha}^\dagger S(H)\vec{\alpha}]\right ]_{\vec{\alpha}_{anc}=0}\ \times e^{-\vec{\alpha}_s^\dagger\cdot\vec{\alpha}_s}
 \end{eqnarray}
 where 
  \begin{equation}
\vec{\alpha}_{anc}=\left (\begin{array}{c}
				\alpha_{N+1}\\
				\alpha_{N+2}\\
				\vdots\\
				\alpha_{N+K}
				\end{array}\right )
\end{equation}
and
  \begin{eqnarray*}
  \partial^*_n & = & \frac{\partial}{\partial \alpha_n^*}\\
    \partial_n & = & \frac{\partial}{\partial \alpha_n}
    \end{eqnarray*}
This expression enables us to construct the relevant Q-symbol for the conditional state transformation by expanding the function ${\cal U}_H(\vec{\alpha})$ as a power series and reading off the relevant coefficient. This function thus plays the role of a Feynman propagator in ordinary quantum mechanics. It describes the many ways in which a classical coherent amplitude can propagate through the linear optical network, keeping track of the manyfold coherent amplitudes for reflection and transmission. This captures in a formal way a heuristic that has often been used to construct conditional gates in the past

As an introduction we consider two modes, $a_1,a_2$, coupled  with a beam splitter interaction, so that the unitary transformation of the modes is given by
\begin{equation}
H=\left (\begin{array}{cc}
		0 & \theta\\
		\theta & 0
		\end{array}\right )
\end{equation}
then the transformation is parameterised by a single parameter, $\theta$ and 
\begin{eqnarray}
a_1(\theta) & = & \cos\theta a_1-\sin\theta a_2\\
a_2(\theta) & = & \cos\theta a_2-\sin\theta a_1
\end{eqnarray}
Then we see that
\begin{equation}
{\cal U}_H(\alpha_1,\alpha_2)=\exp[(|\alpha_1|^2+|\alpha_2|^2)\cos\theta+(\alpha_2^*\alpha_1-\alpha_1^*\alpha_2)\sin\theta]e^{-(|\alpha_1|^2-|\alpha_2|^2)}
\end{equation}
We now assume that photons are counted on mode $a_2$ and calculate the conditional state for mode $a_1$ for the four cases: $n=0,1$ and $m=0,1$.

The  conditional state of mode $a_1$ is given by,
\begin{equation}
|\psi^{(n,m)}\rangle_1 = \frac{1}{\sqrt{p(m|n)}} \hat{E}(n|m)|\psi\rangle_1
\end{equation}
where 
\begin{equation}
p(m|n)=\mbox{}_1\langle \psi|\hat{E}^\dagger(n|m)\hat{E}(n|m)|\psi\rangle_1
\end{equation}

The Q-symbol for the operator $\hat{E}(n|m)$ is
\begin{equation}
{\cal E}(\alpha_1|n,m)=\left (\frac{\partial}{\partial \alpha_2^*}\right )^m\left (\frac{\partial}{\partial\alpha_2}\right )^n\left [\exp[(|\alpha_1|^2+|\alpha_2|^2)\cos\theta+(\alpha_2^*\alpha_1-\alpha_1^*\alpha_2)\sin\theta]\right ]_{\alpha_2=0} e^{-|\alpha_1|^2}
\end{equation}

The corresponding operators are
\begin{eqnarray*}
\hat{E}(0|0) & = & \sum_{n=0}^\infty\frac{(\cos\theta-1)^n}{n!}(a_1^\dagger)^n a_1^n= :\  e^{-ln(\cos\theta) a_1^\dagger a_1} \ :\\
\hat{E}(1|1) & = & \cos\theta \hat{E}(0|0)-\sin^2\theta a_1^\dagger \hat{E}(0|0) a_1\\
\hat{E}(1|0)  & = & -a_1^\dagger \sin\theta \hat{E}(0|0)   \\
\hat{E}(0|1) & = & \sin\theta \hat{E}(0|0) a_1
\end{eqnarray*}

Now consider a three mode example defined by the transformation
\begin{equation}
U^\dagger(H)\vec{a}U(H)=\left (\begin{array}{ccc}
			s_{11} & s_{12} & s_{13}\\
			s_{21} & s_{22} & s_{23}\\
			s_{31} & s_{32} & s_{33}
			\end{array}\right )\left (\begin{array}{c}
							a_1\\
							a_2\\
							a_3\end{array}\right )
\end{equation}
We will  regard $a_3$ as the ancilla mode, prepared in a single photon state $|n\rangle_3$ and we will condition off a count of $"1"$ at this mode after the optical system. Thus we have a three mode example with $n_3=1,m_3=1$ and we wish to calculate the two mode conditional measurement operator $\hat{E}(1|1)$ acting on modes $a_1,a_2$. 

The formalism gives
\begin{equation} 
\hat{E}(1|1)=s_{33}\hat{A}+(s_{13}a_1^\dagger+s_{23}a_2^\dagger)\hat{A}(s_{31}a_1+s_{32}a_2)
\label{three_mode}
\end{equation}
where the Q-symbol for $\hat{A}$ is given by
\begin{equation}
{\cal A}=\exp[-(1-s_{11})|\alpha_1|^2-(1-s_{22})|\alpha_2|^2+s_{12}\alpha_1^*\alpha_2+s_{21}\alpha_1\alpha_2^*]
\end{equation}
Using the theorem\cite{Traux}
\begin{eqnarray*}
\mbox{} & & \langle \alpha_1,\alpha_2|e^{\lambda_1 a_1^\dagger a_2}e^{\mu_1a_1^\dagger a_1+\mu_2 a_2^\dagger a_2} e^{\lambda_2 a_1 a_2^\dagger}|\alpha_1,\alpha_2\rangle\\
& = & \exp\left [-|\alpha_2|^2(1-e^{\mu_2})-|\alpha_1|^2(1-e^{\mu_1}-\lambda_1\lambda_2e^{\mu_2})+(\alpha_1^*\alpha_2\lambda_1+\alpha_1\alpha_2\lambda_2)e^{\mu_2}\right ]
\end{eqnarray*}
we see that
\begin{equation}
\hat{A}=e^{\lambda_1 a_1^\dagger a_2}e^{\mu_1a_1^\dagger a_1+\mu_2 a_2^\dagger a_2} e^{\lambda_2 a_1 a_2^\dagger}
\end{equation}
with 
\begin{eqnarray}
e^{\mu_2} & = & s_{22}\\
e^{\mu_1} & = &  s_{11}-\frac{s_{12}s_{21}}{s_{22}}\\
\lambda_1 & = &\frac{s_{12}}{s_{22}}\\
\lambda_2 & = & \frac{s_{21}}{s_{22}}
\end{eqnarray}





\begin{figure}[h!]
\centering{\includegraphics[scale=0.4]{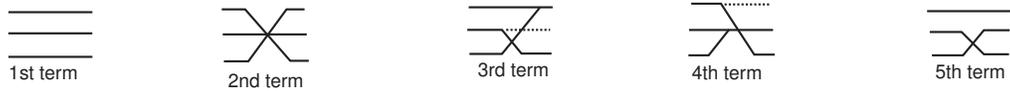}}
\caption{A set of diagrams corresponding to each of the terms in $
s_{33}\hat{A}\ +\ a_1^\dagger s_{13}\hat{A}s_{31}a_1 +a_1^\dagger s_{13}\hat{A}s_{32}a_2+ a_2^\dagger s_{23}\hat{A}s_{31}a_1+ a_2^\dagger s_{23}\hat{A}s_{32}a_2
$. The modes are labeled $1,2,3$ from top to bottom.}
\label{diagram}
\end{figure}

The form  of this Eq.(\ref{three_mode}) suggests an interpretation as a `sum over histories'  for the different ways a photon can go from input to detection. For each of the terms in Eq. (\ref{three_mode}),  we can assign a diagram, see figure\ref{diagram}

As an explicit example, consider the situation where all beam splitters are $50/50$. 
In this case 
\begin{equation}
S=\frac{1}{2\sqrt{2}}\left (\begin{array}{ccc}
				\sqrt{2}+1 & \sqrt{2}-1 & \sqrt{2}\\
				\sqrt{2}-1 & \sqrt{2}+1 & -\sqrt{2}\\
				\sqrt{2} & -\sqrt{2} & -2
				\end{array}\right )
\end{equation}
The 
\begin{equation}
\hat{E}(1|1)=-\frac{1}{\sqrt{2}}\hat{A}+\frac{1}{4}(a_1^\dagger-a_2^\dagger)\hat{A}(a_1-a_2)
\end{equation}
If the input state on modes $a_{1,2}$ is $|\psi\rangle=|1,1\rangle$, the (un normalised) output state is given by
\begin{equation}
\hat{E}(1|1)|1,1\rangle=-\frac{3}{8}(|2,0\rangle+|0,2\rangle)-\frac{1}{4\sqrt{2}}|1,1\rangle
\end{equation}
The norm of this state gives the probability for this event,
\begin{equation}
P(1|1)=5/16
\end{equation}
%




We have presented a method based on quasi probability distribution for constructing conditional linear optical transformations.  The explicit construction involves a power series expansion of the Q-symbol for the linear optical network that defines the process. This method lends itself to a diagrammatic interpretation. More importantly it lends itself to a computer based symbolic computation. We believe this method will assist in finding new conditional state transformations.

 I would like to thank Andrew White and Tim Ralph for useful comments. I acknowledge the support of the Australian Research Council.

\end{document}